\documentclass[preprint,12pt,authoryear]{elsarticle}

\usepackage{amssymb}
\def\la{\hbox{\raise.5ex\hbox{$<$} 
     \kern-1.1em\lower.5ex\hbox{$\sim$}}} 
\def\ga{\hbox{\raise.5ex\hbox{$>$} 
     \kern-1.1em\lower.5ex\hbox{$\sim$}}} 

\journal{Computer Physics Communications}

\begin{document}

\begin{frontmatter}



\title{On the correction of conserved variables for numerical
         RMHD with staggered constrained transport}


\author{Jos\'e-Mar\'{\i}a Mart\'{\i}}

\ead{jose-maria.marti@uv.es}

\address{Departamento de Astronom\'{\i}a y Astrof\'{\i}sica \\
  Universitat de Val\`encia \\ E-46100 Burjassot
  (Val\`encia) \\ SPAIN}

\address{Observatori Astron\`omic \\ Universitat de Val\`encia \\
E-46980 Paterna (Val\`encia) \\ SPAIN}

\begin{abstract}
  Despite the success of the combination of conservative schemes and
staggered constrained transport algorithms in the last fifteen years,
the accurate description of highly magnetized, relativistic flows with
strong shocks represents still a 
challenge in numerical RMHD. The present paper focusses in the 
accuracy and robustness of several correction algorithms for the
conserved variables, which has become a crucial ingredient in the
numerical simulation of problems where the magnetic pressure dominates 
over the thermal pressure by more than two orders of magnitude.

  Two versions of non-relativistic and fully relativistic
corrections have been tested and compared using a magnetized
cylindrical explosion with high magnetization ($ \ga 10^4$) as
test. In the non-relativistic corrections, the total energy is
corrected for the difference in the classical magnetic energy term
between the average of the staggered fields and the conservative ones, 
before (CA1) and after (CA1') recovering the primitive
variables. These corrections are unable to pass the test at any
numerical resolution. The two relativistic approaches (CA2 and CA2'),
correcting also the magnetic terms depending on the flow speed in both
the momentum and the total energy, reveal as much more robust. These
algorithms pass the test succesfully and with very small deviations of
the energy conservation ($\la 10^{-4}$), and very low values of the
total momentum ($\la 10^{-8}$). In particular, the algorithm CA2'
(that corrects the conserved variables after recovering the primitive
variables) passes the test at all resolutions. 

  The numerical code used to run all the test cases is briefly described.
\end{abstract}


\end{frontmatter}



\section{Introduction}
\label{s:intro}

The popularization of the use of conservative methods in numerical
RMHD, introduced in the late nineties \citep{KS99,Ko99,Ba01}, has
revolutionized the research in several fields of Relativistic
Astrophysics as, e.g., the study of the jets emanating from AGN
\citep{Ko99a,LA05,KM08,MR10,ML12}, the structure and dynamics of
pulsar wind nebulae \citep{KL03,DA04,PK13}, or the production of GRB
\citep{KV09,KV10,RG11}. 

To prevent the generation of artificial forces that can falsify the
solutions, conservative schemes in numerical RMHD must be supplemented 
with some additional procedure  to keep the magnetic field solenoidal
along the simulation. Among the most succesful strategies is the
combination of numerical algorithms in conservation form to evolve
cell-centered (finite-difference, finite-volume) representations of
the hydrodynamical variables, and constrained transport (CT)
algorithms for the magnetic field discretized on a staggered grid. The
CT scheme, based in the conservation of magnetic flux across closed
surfaces, maintains the value of the divergence of the magnetic
field to the accuracy of machine round-off errors on a specific
discretization. It was originally developed for artificial viscosity methods
\citep{EH88}. \cite{DW98}, \cite{RM98}, and \cite{BS99} combined the
CT discretization with conservative schemes. \cite{LD00,LD04}
  developed the upwind constrained transport (UCT) strategy, which
  extends  the CT method to high-order upwind schemes. At present,
many RMHD codes
\citep{Ko99,DB03,LA05,SS05,AZ06,MB06,MB07,GR07,DZ07,EL10,BS11} use
this approach. 

The combination of conservative algorithms and staggered CT schemes  
has proven to be very succesful in the simulation of highly 
magnetized, relativistic flows with strong shocks. However the 
existence of two sets of variables defined on different grids requires 
in the most extreme cases some additional work to make the 
conservative and CT steps fully consistent. The proposed 
solutions \citep{Ko99,BS99,To00,MB06} rely on correcting the conserved 
quantities after each time step to make them consistent with the 
staggered fields. The present paper explores the performance of
several correction algorithms focussing in their accuracy and
robustness.  

The evolution system guarantees the fulfillment of this constraint for
an initially divergence-free magnetic field at all later times, but to
satisfy the constraint in numerical simulations of MHD flows poses a
challenge.

\section{Conservative methods for RMHD and Constrained Transport}
\label{s:HRSC-CT}

\subsection{The RMHD equations in conservation form}
\label{ss:conservation}

The equations of ideal RMHD, representing the conservation of
rest-mass, momentum and energy of the magnetized fluid together with
the induction equation can be written as a system of conservation
laws\footnote{The solenoidal condition for the magnetic field is
  guaranteed by the evolution system however it is by no means trivial
  to satisfy numerically.}. In Minkowski spacetime and Cartesian coordinates
($\{i,j,k\}=\{x,y,z\}$) this system reads\footnote{Throughout this
  paper, besides using units in which the speed of light is set to unity, a factor $\sqrt{4
\pi}$ is absorbed in the definition of the magnetic field.}
\begin{equation}
  \frac{\partial {\bf U}}{\partial t} +
  \frac{\partial {\bf F}^i({\bf U})}{\partial x^i}=0 \, ,
\label{21}
\end{equation}
where the vector of conserved variables, ${\bf U}$, and the fluxes, ${\bf F}^i$, are
the column vectors,
\begin{eqnarray}
{\bf U} & = & \left(\begin{array}{c}
  D    \\
  S^j  \\
  E \\
  B^k
\end{array}\right),
\label{state_vector}
\end{eqnarray}
\begin{eqnarray}
{\mathbf F}^i & =& \left(\begin{array}{c}
  D v^i \\
  S^j v^i + p^{*} \delta^{ij} - b^j B^i/W \\
  E v^i + p^{*} v^i - b^0 B^i/W \\
  v^i B^k - v^k B^i
\end{array}\right).
\label{flux2}
\end{eqnarray}
In these equations, $D$, $S^j$, $E$ and $B^k$ are the rest-mass density,
the momentum density of the magnetized fluid in $j$-direction, the
total energy density, and the magnetic field measured in the laboratory frame,
\begin{equation}
\label{eq:D}
  D = \rho W,
\end{equation}
\begin{equation}
\label{eq:Sj}
  S^j = \rho h^* W^2 v^j - b^0 b^j,
\end{equation}
\begin{equation}
\label{eq:tau}
  E = \rho h^* W^2 - p^* - (b^0)^2,
\end{equation}
\noindent
where $\rho$ is the proper rest--mass density of the fluid, $p^*$ is
the total pressure, and $h^*$ is the specific enthalpy including the
contribution of the magnetic field. These two last quantities are
defined according to  
\begin{equation}
\label{eq:p*}
  p^* = p + \frac{b^2}{2}
\end{equation}
\begin{equation}
\label{eq:h*}
  h^* = 1 + \varepsilon + p/\rho + b^2/\rho,
\end{equation}
where $p$ is the fluid pressure and $\varepsilon$ its specific
internal energy. $b^\mu$ ($\mu =
0,1,2,3$) are the components of the 4-vector representing the magnetic
field in the fluid rest frame and $b^2$ stands for $b^\mu b_\mu$,
where summation over repeated indices is assumed. $v^i$ are the components of the fluid
3-velocity in the laboratory frame, which are related to the flow
Lorentz factor, $W$, according to: 

\begin{equation}
\label{eq:W}
  W = \frac{1}{\sqrt{1-v^i v_i}}.
\end{equation}

The following relations hold between the components of the
magnetic field 4-vector in the comoving frame and the three vector
components $B^i$ measured in the laboratory frame:
\begin{eqnarray}
\label{b0}
  b^0 & = & W\, {\bf B} \cdot {\bf v} \ , \\
  \label{bi}
  b^i & = & \frac{B^i}{W} + b^0 v^i \ ,
\end{eqnarray}
where ${\bf v}$ and ${\bf B}$ denote the 3-vectors $(v^x,v^y,v^z)$ and
$(B^x,B^y,B^z)$, respectively. The square of the modulus of the
magnetic field can be written as
\begin{equation}
  b^2 = \frac{{B}^2}{W^2} + ({\bf B} \cdot {\bf v})^2 
\end{equation}
with $B^2 = B^iB_i$.

An equation of state that relates the thermodynamic
variables, e.g., $p = p(\rho, \varepsilon)$, is needed to close
the system.

To make more clear the corrections to be performed on them, the
momentum and total energy densities are written making explicit their
dependence on the magnetic field in the laboratory frame, ${\bf B}$
\begin{equation}
\label{eq:Sj}
  S^j = (\rho h W^2 + B^2) v^j - ({\bf v} \cdot {\bf B}) B^j,
\end{equation}
\begin{equation}
\label{eq:tau}
  E = \rho h W^2 - p + \frac{{B}^2}{2} + \frac{v^2 B^2 -  ({\bf v}
    \cdot {\bf B})^2}{2}
\end{equation}
\noindent
($v^2 = v^i v_i$, $h = 1 + \varepsilon + p/\rho$).

\subsection{Conservative methods for RMHD and Constrained Transport}
\label{ss:HRSC-CT}

 Conservative methods exploit the conservation properties of the
system of equations and can be directly applied to solve the equations
of ideal RMHD written in conservation form, Eq.~(\ref{21}).

  In these methods, 
\begin{equation}
\frac{d {\bf U}_{i,j,k}}{dt} = {\cal L}^x_{i,j,k} + {\cal
  L}^y_{i,j,k} + {\cal L}^z_{i,j,k},
\label{eq:cscheme}
\end{equation}
where ${\bf U}_{i,j,k}$ is the value of the conserved variable ${\bf U}$
at the point $(x_i,y_j,z_k)$ (finite-difference approach) or its
volume average at the cell centered in that point (finite-volume
approach), and
\begin{equation}
{\cal L}^x_{i,j,k} = - \frac{1}{\Delta x_i} \left( \hat{{\bf
    F}}^x_{i+1/2,j,k} - \hat{{\bf
    F}}^x_{i-1/2,j,k} \right)
\end{equation}
(and similar expressions for ${\cal L}^y_{i,j,k}$ and ${\cal
  L}^z_{i,j,k}$). Quantities $\hat{{\bf F}}^x_{i+1/2,j,k}$ are the
numerical fluxes. In the Godunov-type methods (finite-volume
approach), these fluxes are obtained from the solution of Riemann 
problems at cell interfaces $(x_{i+1/2},y_j,z_k)$, where the initial
left and right states, ${\bf U}_{L,i+1/2}$, ${\bf U}_{R,i+1/2}$
are reconstructed values from the corresponding cell averages.

  The algorithm defined in Eq.~(\ref{eq:cscheme}) can be used to evolve
system~(\ref{21}) in time. However, additional care has to be taken to
prevent the divergence of the magnetic field to grow with time. Staggered CT
algorithms maintain the divergence of ${\bf B}$ exactly (i.e., to the
accuracy of machine round-off errors) in each numerical cell. In this
approach, the normal component of the surface-centered magnetic field
on the cell interface at $(x_{i+1/2},y_j,z_k)$ is evolved according to
a discretized version of the induction equation
\begin{eqnarray}
\nonumber
  \frac{d B^{x}_{i+1/2, j,k}}{dt}  & = & \frac{1}{\Delta y}
  \big( \hat{\Omega}^{z}_{i+1/2, j+1/2,k} - \hat{\Omega}^{z}_{i+1/2, j-1/2,k}
  \big) - \\
   & & \frac{1}{\Delta z}
  \big( \hat{\Omega}^{y}_{i+1/2, j,k+1/2} - \hat{\Omega}^{y}_{i+1/2, j,k-1/2} \big),
\end{eqnarray}
where quantities $\hat{\Omega}^{i}$ are discretized representations of
the components of ${\bf \Omega} = {\bf v} \times {\bf B}$ computed at
cell edges in terms of spatial and temporal interpolations of the
magnetic field and the velocity, or the numerical fluxes of the
conservative step.

  Once the staggered magnetic fields have been computed, the
corresponding cell-centered fields can be obtained by
interpolation. For second order accuracy, a linear interpolation is
enough and
\begin{equation}
  B^{x}_{i, j,k} = \frac{1}{2} (B^{x}_{i-1/2, j,k} + B^{x}_{i+1/2, j,k})
\label{Bxaverage}
\end{equation}
(and similar expressions for $B^{y}_{i, j,k}$ and $B^{z}_{i, j,k}$).

  In its simplest version, the evolution scheme consists of a
conservative step in which the rest-mass, momentum and energy
densities are advanced in time, and a CT step to advance the staggered
magnetic fields and obtain the cell-centered ones. Finally, inherent
to all the conservative methods in RMHD is to solve an implicit
algebraic system to recover the primitive variables (needed to compute
the numerical fluxes), ${\bf V} = (\rho,p,v^i,B^k)$ from the conserved
ones, ${\bf U}$, at every time step. 

\subsection{Setting the problem}
\label{ss:problem}

  According to the previous algorithm, at time $t = t^{n+1}$, the
conserved variables have been advanced consistently with the magnetic
fields defined at cell centers in the previous time step and are hence
consistent with a magnetic field at cell centers, $B^{x,n+1}_{{\rm cell}, i, j,k}$,
$B^{y,n+1}_{{\rm cell}, i, j,k}$ as computed with the cell-centered
scheme (conservative step). However these fields are different from
those defined at cell centers, $B^{x,n+1}_{{\rm stag}, i, j,k}$,
$B^{y,n+1}_{{\rm stag}, i,  j,k}$, as the average of the staggered
fields using Eq.~(\ref{Bxaverage}). The fact that these two sets of magnetic
fields are different is on the basis of the inconsistency of the
algorithm, which makes it unsuitable for problems where the magnetic
pressure dominates over the thermal pressure by more than two orders
of magnitude.

  The proposed solutions \citep{Ko99,BS99,To00,MB06} rely on
redefining the conserved quantities after each time step to make them
consistent with the staggered field. 

\section{Correction of the conserved variables}
\label{s:correction}

  Relying in the approach proposed by \cite{BS99} for classical MHD,
\cite{MB06} proposed a non-relativistic correction of the conserved
energy after each time step according to
\begin{equation}
E_{\rm stag} = E_{\rm cell} - \frac{{B}^2_{\rm cell} - {B}^2_{\rm stag}}{2}.
\end{equation}
In the previous expression, ${\bf B}_{\rm cell}$ and ${\bf B}_{\rm
  stag}$ are, respectively, the magnetic fields at some cell as
computed in the 
conservative step, and by averaging the staggered fields. $E_{\rm cell}$
and $E_{\rm stag}$ are the corresponding conserved energy densities.

  The procedure is computationally efficient since it does not involve
extra calls to the primitive recovery procedure per time step, although
as the rest of correction algorithms proposed, it forces to evolve the
two equations of the magnetic field components normal to the sweep in
the conservative step\footnote{Since all the procedures are based
  on the advance in time of the cell-centered magnetic fields with the
  conservative algorithm, the number of three-dimensional variables
  increases in three. However, besides the additional calls to the primitive
  recovery procedure, where appropriate, the only extra computational
  cost is the pure advance of the two magnetic field components normal
  to the direction of the sweep, since these components are also
  reconstructed and used in the solution of the Riemann problems 
  even in the case of no correction.}. On the other hand, the
non-relativistic nature of the correction dismisses all the magnetic
terms depending on the flow velocity in the momentum and energy
densities, making this correction questionable in those cases
combining highly relativistic flow velocities and large magnetic
fields. 

  This correction can be applied in two ways depending whether the
correction of the energy is done before (correction algorithm CA1) or
after (CA1') recovering the primitive variables. A potential drawback
of the second approach is that hydrodynamic, primitive variables are
not made consistent with the staggered fields. 

  In this work, we propose a full-relativistic correction in which the
non-relativistic procedure CA1 is used as a first approximation. The
correction (CA2) proceeds as follows: 

\begin{enumerate}

\item Obtain an approximation to the total energy consistent with the
  staggered fields:
\begin{equation}
E^{(1)}_{\rm stag} = E_{\rm cell} - \frac{{B}^2_{\rm cell} - {B}^2_{\rm stag}}{2}.
\end{equation}
\item  Obtain an approximation to the primitive variables, namely
  ${\bf V}^{(1)}$, from $\{D, {\bf S}, E^{(1)}_{\rm stag}, {\bf B}_{\rm stag} \}$.

\item Use the flow velocity ${\bf v}^{(1)}$ to complete the relativistic
  correction of the momentum and energy densities:
\begin{equation}
S^{i\,(2)}_{\rm stag} = S^i - ({B}^2_{\rm cell} - {B}^2_{\rm
  stag}) v^{i\,(1)} + {\bf v}^{(1)} \cdot (B^i_{\rm cell} {\bf B}_{\rm cell} - B^i_{\rm stag} {\bf B}_{\rm stag})
\end{equation}
\begin{equation}
E^{(2)}_{\rm stag} = E^{(1)}_{\rm stag} - \frac{({v}^{(1)})^2}{2} ({B}^2_{\rm cell} - {B}^2_{\rm
  stag}) + \frac{({\bf v}^{(1)} \cdot {\bf B}_{\rm
    cell})^2 - ({\bf v}^{(1)} \cdot {\bf B}_{\rm
    stag})^2}{2}
\end{equation}

\item Obtain the primitive variables from $\{D, {\bf S}^{(2)}_{\rm
    stag}, E^{(2)}_{\rm stag}, {\bf B}_{\rm stag} \}$.

\end{enumerate}

  Steps 1. and 2. correspond to the classical correction CA1. Steps
3. and 4. complete the relativistic correction. The procedure
involves one extra call to the recovery of primitive variables. 
 
  Finally, as in the correction algorithm CA1', the recovery of the
primitive variables can be done prior to the correction of the
conserved variables, leading to a very simple and computationally
efficient algorithm (CA2'). In this case, the primitive variables are
recovered using the cell-centered magnetic fields as advanced by the
conservative algorithm, ${\bf B}_{\rm cell}$, and then the conserved
variables recalculated from these primitive variables and ${\bf
  B}_{\rm stag}$. This is the approach followed by \cite{Ko99}. As in
the case of correction CA1', a potential drawback of this approach is
that hydrodynamic, primitive variables are not made consistent with
the staggered fields (but, in contrast to CA1', the corrected
  conserved variables do include the relativistic corrections).

\section{Numerical tests}
\label{s:tests}

\subsection{The numerical code}
\label{ss:code}

  The basic ingredientes of the code used to test the correction
procedure are the following:

\begin{itemize}

\item[i)] Cell reconstruction: second-order accurate values of
  the primitive variables ${\bf V}$ at the left and
  right ends of the cells are obtained with linear functions and
  several limiters (MINMOD, VAN LEER, MC). MC and VAN LEER limiters
  can be degraded to MINMOD in case of strong shocks. No jump is
  allowed in the normal component of ${\bf B}$ at a cell boundary and
  the corresponding staggered magnetic field is used.

\item[ii)] Riemann solvers: intercell numerical fluxes are computed by
  means of HLL and HLLC \citep{MB06} Riemann solvers. Accurate bounds
  of the maximum speeds of left and right propagating waves are obtained by
  solving the corresponding characteristic equation for the left and
  right states of each numerical interface.

\item[iii)] Time advance: the multidimensional equations of RMHD are
  advanced in time in an unsplit manner using TVD-preserving
  Runge-Kutta methods of second and third order \citep{SO88,SO89}. The
  time step is determined according to $\displaystyle{\Delta t = \frac{C\!F\!L}{\sqrt{2}}
  \times \min_{i,j}\left(\frac{\Delta x}{|\lambda_{x,i,j}|},\frac{\Delta
    y}{|\lambda_{y,i,j}|}\right)}$ (2D, planar symmetry version), where $\lambda_{x,i,j}$ and
$\lambda_{y,i,j}$ are the speeds of the fastest waves propagating in
cell $i,j$ along the $x$ and $y$ direction, respectively. 

\item[iv)] Constrained transport scheme as in \cite{BS99}.

\item[v)] Primitive variables are recovered as in the 1D$_W$ method of
  \cite{NG06} and solving the resulting equation in $Z = \rho h W^2$
  by bisection.

\end{itemize}

  The code advances the total energy density without the rest-mass 
energy density, i.e., $E - D$. This strategy improves the performance
of the conservative scheme when the total energy is dominated by the
rest-mass energy. However, in this case it does not produce any
effect, since in the selected test (see next Section) the total energy
density is larger than the rest-mass energy density by several orders
of magnitude.

\subsection{The test: cylindrical magnetized blast wave}
\label{ss:smbw}

  The setup for this test is taken from \cite{Ko99}. A cylindrical
region of radius $r=0.8$ with density $\rho = 10^{-2}$ and thermal
pressure $p = 1$ is embedded in a static uniform medium
with $\rho = 10^{-4}$ and $p = 3 \times10^{-5}$. A linear smoothing function
is applied for $0.8 < r < 1$. The whole region is threaded by a
constant horizontal magnetic field in the $x$-direction, $B_x = 1$. An ideal gas
equation of state with $\gamma = 4/3$ is used. 

  The difference in pressure between the cylindrical region and the
ambient medium produces the expansion of the central region delimited
by a fast forward shock propagating radially at almost the speed of
light. Because of the strong sideways magnetic confinement
an elongated structure develops in the $x$ direction with a maximum
Lorentz factor of $W \simeq 4.5$. This problem is particularly
challenging because of the very large magnetization $\beta = b^2/(2p)
= 1.67 \times 10^4$.

  The test is solved in Cartesian $(x,y)$ coordinates. The numerical
grid covers a $[-6,6] \times [-6,6]$ square with the center of the
cylindrical region at $(0,0)$. Open boundary conditions are
placed along the boundaries of the computational domain.

\begin{figure}
\includegraphics[width=14.cm,angle=0]{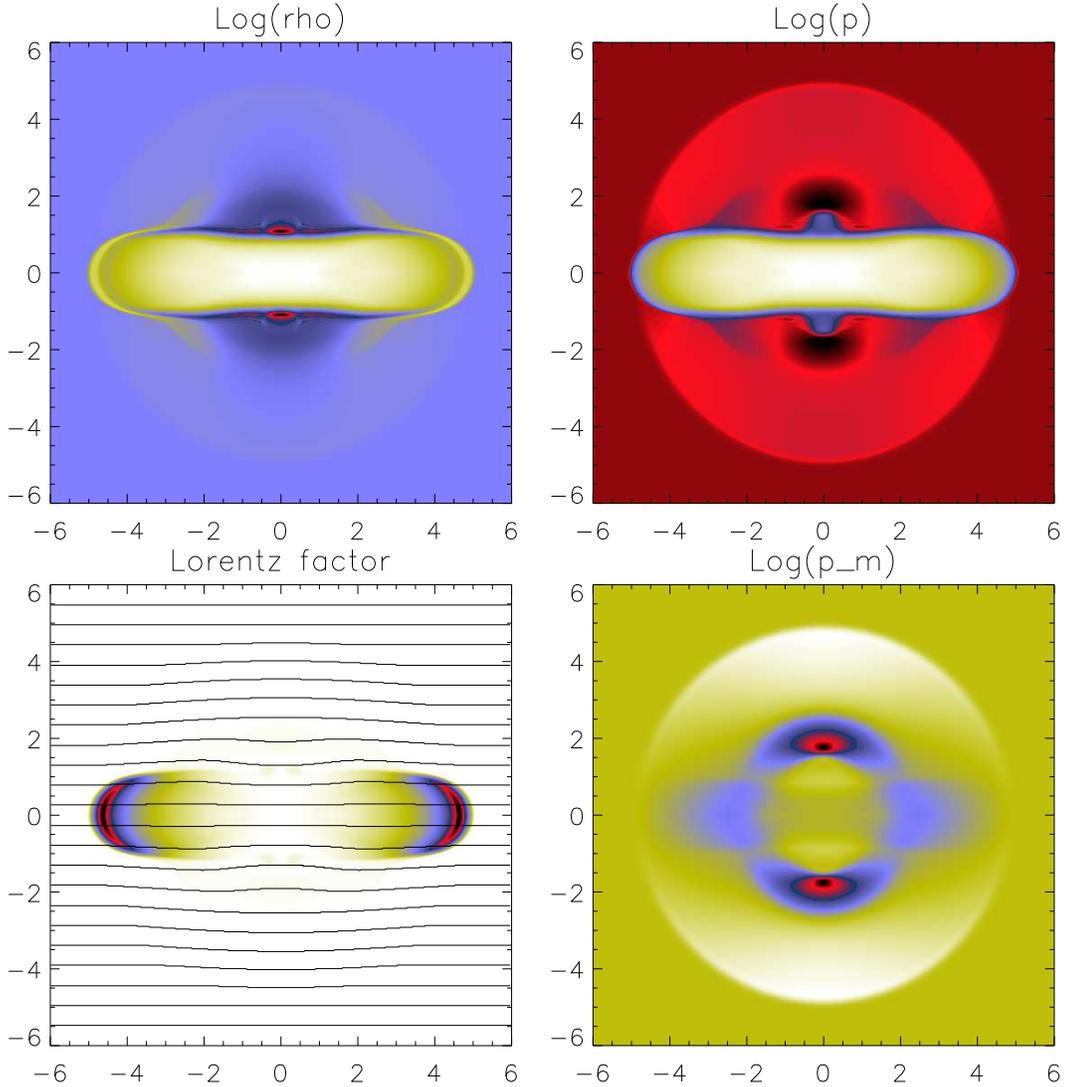}
\caption{Proper rest-mass density, gas pressure and magnetic
pressure (in logarithmic scale), and flow Lorentz factor at $t = 4.0$
as computed in the reference run of the cylindrical magnetized blast
wave test discussed in the text. The magnetic field lines are 
printed on top of the Lorentz factor plot. $\log \rho \in
[-5.30,-2.71]$, $\log p \in [-4.65,-0.94]$, $\log p_m \in
[-0.60,-0.20]$, $W \in [1.00,4.63]$.}
\label{f1}
\end{figure}

  A reference model using the HLLC Riemann solver and the
third order Runge-Kutta (RK3) for time advance has been ran. For the cell
reconstrucion we have used the VAN LEER limiter degraded to MINMOD (VLMM)
when the relative jump in thermal or magnetic pressure within a shock
exceeds $0.5$. The relativistic correction proposed in this paper
(CA2) was chosen to correct the conserved variables after each time
substep. A $C\!F\!L$ of $0.45$ and a numerical resolution of $N_x \times N_y
= 512^2$ cells complete the initial setup. Figure~\ref{f1} shows the
distributions of proper rest-mass density, gas pressure and magnetic
pressure (in logarithmic scale), and flow Lorentz factor at $t = 4.0$
as computed in the reference run. The fast magnetosonic shock and the
elongated horizontal structure are clearly seen. 

%%%%%%%%%%%%%%%%%%%%%%%%%%%%%%%%%%%%%%%%%%%%%%%%%%%%%%%%%%%
%
%
%%%%%%%%%%%%%%%%%%%%%%%%%%%%%%%%%%%%%%%%%%%%%%%%%%%%%%%%%%%

%%%%%%%%%%%%%%%%%%%%%%%%%%%%%%%%%%%%%%%%%%%%%%%%%%%%%%%%%%%
%
\begin{table}[]
\begin{center}
\caption{Total momentum (first row in two-row table entries) and
  relative change in total energy (secod row) in the
  cylindrical magnetized blast wave test discussed in the text at $t = 4.0$, as a
  function of the correction algorithm and numerical resolution. Table
entries with a single number display the time at which the run stopped
abnormally.}
\label{t1}

\bigskip

\begin{tabular}{ccclcc}
\hline\\[-2.ex]
 $N_x \times N_y$ & CA0 & CA1 & CA1' & CA2 & CA2' \\[0.6ex]
\hline\\[-2.ex]
$128^2$ & $2.66 \times 10^{-2}$ & $1.13$ & $1.18$ & $1.19 \times 10^{-11}$ &
$8.00 \times 10^{-12}$ \\
 &  &  & & $1.51 \times 10^{-4}$ &
$1.55 \times 10^{-4}$ \\
\hline\\[-2.ex]
$256^2$ & $2.29 \times 10^{-2}$ & $0.70$ & $0.73$ & $2.17 \times 10^{-11}$ &
$7.50 \times 10^{-10}$ \\
 &  &  & & $7.69 \times 10^{-5}$ &
$7.91 \times 10^{-5}$ \\
\hline\\[-2.ex]
$512^2$ & $2.72 \times 10^{-2}$ & $0.39$ & $0.42$ & $4.11 \times 10^{-11}$ &
$1.55 \times 10^{-8}$ \\
 &  &  & & $2.60 \times 10^{-5}$ &
$2.70 \times 10^{-5}$ \\
\hline\\[-2.ex]
$1024^2$ & $3.75 \times 10^{-2}$ & $0.31$ & $0.37$ & $3.43$ &
$7.94 \times 10^{-9}$ \\
 &  &  & &  &
$8.08 \times 10^{-6}$ \\
\hline\\

\end{tabular}

\end{center}

\end{table}
%%%%%%%%%%%%%%%%%%%%%%%%%%%%%%%%%%%%%%%%%%%%%%%%%%%%

Table~\ref{t1} displays the total momentum and the relative change in
total energy at $t=4.0$ for the different correction algorithms and
several numerical resolutions from $128^2$ to $1024^2$ cells. For the
reference model (CA2, $N_x \times N_y = 512^2$), the relative change
in total energy is $2.60 \times 10^{-5}$, whereas the total momentum
(initially equal to zero) reaches a value of $4.11 \times
10^{-11}$. The same run without energy nor momentum corrections (CA0) 
crashes at the sixth iteration ($t \simeq 0.03$), whereas the run
using the classical correction (CA1) crashes at $t \simeq 0.39$. The
same run using \cite{Ko99}'s recipe (CA2'), completes the test normally
and with a relative change in total energy of $2.70 \times 10^{-5}$
and a total momentum of $1.55 \times 10^{-8}$. The parameters chosen 
(VLMM,HLLC,RK3) produce a version of the code with very low
dissipation, which makes the algorithms without correction algorithm (CA0)
and with the classical correction (CA1) to fail in passing the test at
any resolution. Code versions using corrections CA2 and CA2' pass the
test succesfully and with very small deviations of the energy
conservation ($\la 10^{-4}$), and very low values of the total
momentum ($\la 10^{-8}$). In the case of the relativistic correction CA2,
the run stopped abnormally for the largest resolution ($1024^2$) close
to the end of the test, at $t=3.43$.  

%%%%%%%%%%%%%%%%%%%%%%%%%%%%%%%%%%%%%%%%%%%%%%%%%%%%%
\begin{table}[]
\begin{center}
\caption{Maximum and average relative differences in thermal
  pressure for the cases using correction algorithm CA2 and CA2', in the
  cylindrical magnetized blast wave test discussed in the text at $t =
  4.0$. The maximum relative difference is computed as
  $\displaystyle{{\rm max}_{i,j} \left\{\frac{|p_{\rm CA2}(i,j) - p_{\rm
          CA2'}(i,j)|}{p_{\rm CA2}(i,j)}\right\}}$. The average
  relative difference is computed as $\displaystyle{\frac{\sum_{i,j} |p_{\rm CA2}(i,j) - p_{\rm
          CA2'}(i,j)|}{\sum_{i,j} p_{\rm CA2}(i,j)}}$. In these
    expressions, $i,j$ span the whole numerical grid.}
\label{t2}

\bigskip

\begin{tabular}{ccc}
\hline\\[-2.ex]
 $N_x \times N_y$ & max & aver  \\[0.6ex]
\hline\\[-2.ex]
$128^2$ & $1.32 \times 10^{-2}$ & $5.53 \times 10^{-5}$ \\
\hline\\[-2.ex]
$256^2$ & $2.77 \times 10^{-2}$ & $2.75 \times 10^{-5}$  \\
\hline\\[-2.ex]
$512^2$ & $3.65 \times 10^{-2}$ & $8.65 \times 10^{-6}$  \\

\hline\\

\end{tabular}

\end{center}

\end{table}
%%%%%%%%%%%%%%%%%%%%%%%%%%%%%%%%%%%%%%%%%%%%%%%%%%%%

  In flows with high magnetization, accuracy problems can lead to
unphysical internal energies and pressures. The correction
algorithm CA2' can be affected by this failure since in this case the
recovery of primitive variables is done before the correction of the
conserved variables and are then obtained from cell-centered
representations of the magnetic field as advanced in the conservative
step. In Table~\ref{t2}, the maximum and average relative
  differences in thermal pressure at $t = 4.0$ for the cases using the
correction algorithms CA2 and CA2' and numerical resolutions $N_x
\times N_y = 128^2, 256^2, 512^2$, are shown.  The values of the
average relative pressure difference between both algorithms tend to
zero with numerical resolution, whereas the maximum relative
difference seems to increase with numerical resolution although
keeping small values ($\approx 1-3\%$). The conclusion is that the
pressure values obtained with the algorithm CA2' are consistent with
those obtained with the algorithm CA2, in which the primitive
variables are recovered once the conserved variables have been
corrected. 

\begin{figure}
\includegraphics[width=14.cm,angle=0]{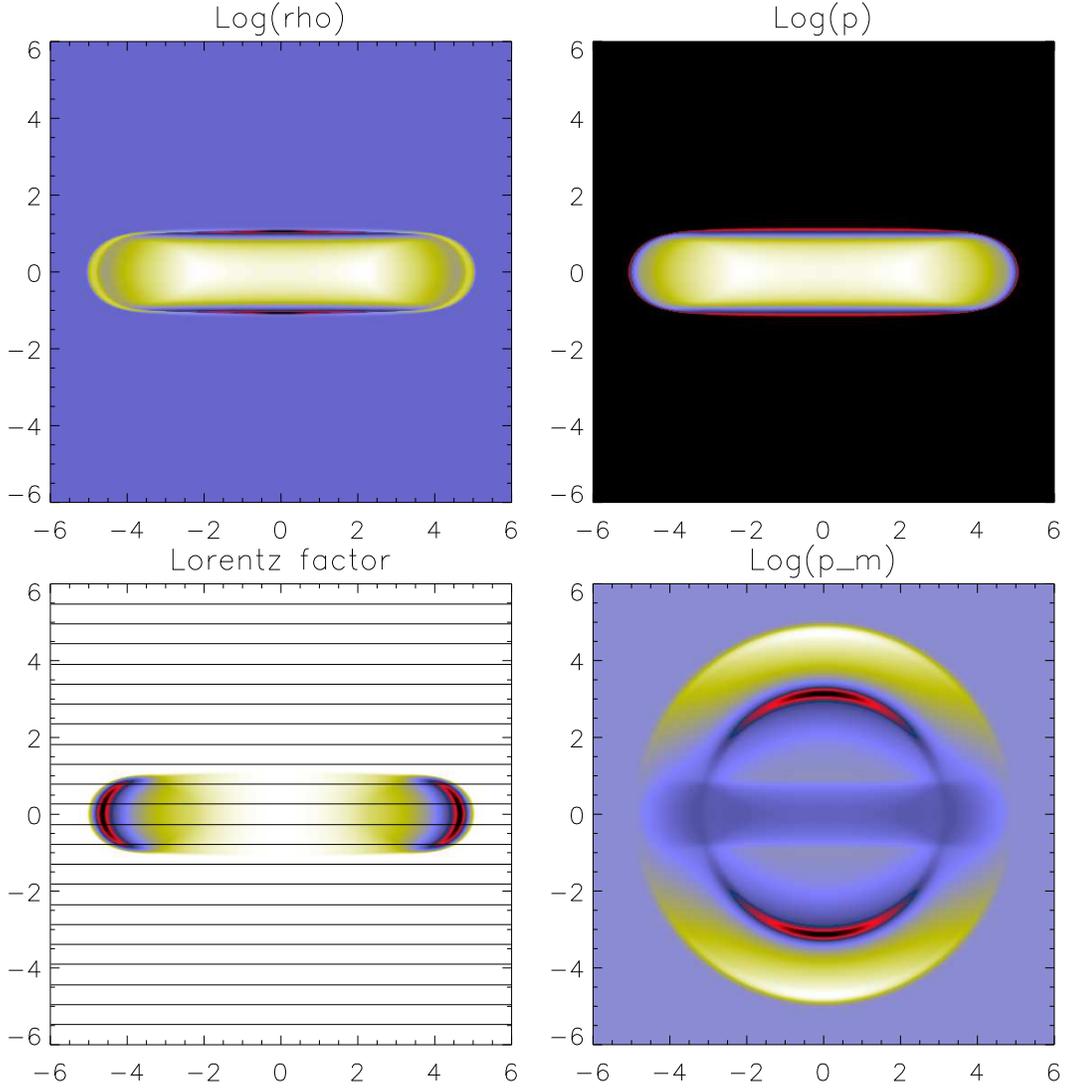}
\caption{Proper rest-mass density, gas pressure and magnetic
pressure (in logarithmic scale), and flow Lorentz factor at $t = 4.0$
as computed in the cylindrical magnetized blast wave test with $B_x =
100$ (magnetization $1.67 \times 10^8$) using the same algorithm
choice as in the test run (VLMM,HLLC,RK3), correction algorithm CA2
and a spatial resolution of $512^2$ cells. The magnetic field lines are  
printed on top of the Lorentz factor plot. $\log \rho \in
[-4.83,-2.75]$, $\log p \in [-4.52,-1.00]$, $\log p_m \in
[3.698950,3.698986]$, $W \in [1.00,4.68]$.}
\label{f2}
\end{figure}

  The same test with $B_x = 10, 100$ (magnetizations $1.67 \times
10^6$, $1.67 \times 10^8$, respectively) has been run without
problems with the same algorithm choice (VLMM,HLLC,RK3) and
correction algorithms  CA2 and CA2' with a spatial resolution of
$512^2$ cells (see Fig.~\ref{f2}). 

%%%%%%%%%%%%%%%%%%%%%%%%%%%%%%%%%%%%%%%%%%%%%%%%%%%%%%%%%%%
%
%
%%%%%%%%%%%%%%%%%%%%%%%%%%%%%%%%%%%%%%%%%%%%%%%%%%%%%%%%%%%

\section{Summary and conclusions}
\label{s:concl}

  Despite the success of the combination of conservative schemes and
staggered constrained transport algorithms in the last fifteen years,
the accurate description of highly magnetized, relativistic flows with
strong shocks flows represents still a 
challenge in numerical RMHD. The present paper focusses in the 
accuracy and robustness of several correction algorithms for the
conserved variables, which has become a crucial ingredient in the
numerical simulation of problems where the magnetic pressure dominates 
over the thermal pressure by more than two orders of magnitude.

  Two versions of non-relativistic and fully relativistic
corrections have been tested and compared using a magnetized
cylindrical explosion with high magnetization ($ \ga 10^4$) as
test. In the non-relativistic corrections, the total energy is
corrected for the difference in the classical magnetic energy term
between the average of the staggered fields and the conservative ones, 
before (CA1) and after (CA1') recovering the primitive
variables. These corrections are unable to pass the test at any 
numerical resolution. The two relativistic approaches (CA2 and CA2'),
correcting also the magnetic terms depending on the flow speed in both
the momentum and the total energy, reveal as much more robust. These
algorithms pass the test succesfully and with very small deviations of
the energy conservation ($\la 10^{-4}$), and very low values of the
total momentum ($\la 10^{-8}$). In particular, the algorithm CA2'
(that corrects the conserved variables after recovering the primitive
variables) passes the test at all resolutions. 

  The numerical code used to run all the test cases is briefly described.

\vspace{5mm}
\noindent
{\it Acknowledgements.}
J.-M. M. acknowledges financial support from the Spanish Ministerio de
Econom\'{\i}a y Competitividad (grants AYA2013-40979-P, and
AYA2013-48226-C3-2-P). The author also ackowledges the referee, L. Del
Zanna, for his comments, which have contributed to improve the first
version of the manuscript.




\begin{thebibliography}{00}


\bibitem[Ant\'on et al.(2006)]{AZ06} Ant\'on, L., Zanotti, O.,
  Miralles, J. A., Mart\'{\i}, J. M., Ib\'a\~nez, J. M., Font,
  J. A. \& Pons, J. A. (2006), ApJ, {\bf 637}, 296
\bibitem[Balsara(2001)]{Ba01} Balsara, D.S. (2001), ApJS, {\bf 132}, 83
\bibitem[Balsara \& Spicer(1999)]{BS99} Balsara, D.S. \& Spicer, S.D. (1999), JCP, {\bf
  149}, 270
\bibitem[Beckwith \& Stone(2011)]{BS11} Beckwith, K. \& Stone,
  J. M. (2011), ApJS, {\bf 193}, article id. 6
\bibitem[Dai \& Woodward(1998)]{DW98} Dai, W. \&  Woodward,
  P. R. (1998), ApJ, {\bf 494}, 317
\bibitem[Del Zanna et al.(2003)]{DB03} Del Zanna, L., Bucciantini,
  N. \& Londrillo, P. (2003), A\&A, {\bf 400}, 397
\bibitem[Del Zanna et al.(2004)]{DA04} Del Zanna, L., Amato, E. \&
  Bucciantini, N. (2004), A\&A, {\bf 421}, 1063
\bibitem[Del Zanna et al.(2007)]{DZ07} Del Zanna, L., Zanotti, O.,
  Bucciantini, N. \& Londrillo, P. (2007), A\&A, {\bf 473}, 11
\bibitem[Etienne et al.(2010)]{EL10} Etienne, Z. B., Liu, Y. T. \&
  Shapiro, S. L. (2010), PRD, {\bf 82}, id. 084031
\bibitem[Evans \& Hawley(1988)]{EH88} Evans, C. R. \& Hawley,
  J. F. (1988), ApJ, {\bf 332}, 659
\bibitem[Giacomazzo \& Rezzolla(2007)]{GR07} Giacomazzo, B. \&
  Rezzolla, L. (2007), CQG, {\bf 24}, S235-S258
\bibitem[Keppens et al.(2008)]{KM08} Keppens, R., Meliani, Z., van
  der Holst, B. \& Casse, F. (2008), A\&A, {\bf 486}, 663
\bibitem[Koide et al.(1999)]{KS99} Koide, S., Shibata, K. \& Kudoh,
  T. (1999), ApJ, {\bf 522}, 727
\bibitem[Komissarov(1999a)]{Ko99} Komissarov, S.S. (1999a), MNRAS, {\bf
  303}, 343
\bibitem[Komissarov(1999b)]{Ko99a} Komissarov, S.S. (1999b), MNRAS,
  {\bf 308}, 1069
\bibitem[Komissarov \& Lyubarsky(2003)]{KL03} Komissarov, S. S.\&
  Lyubarsky, Y. E. (2003), MNRAS, {\bf 344}, L93
\bibitem[Komissarov et al.(2010)]{KV10} Komissarov, S. S., Vlahakis,
  N. \& K\"onigl, A. (2010), MNRAS, {\bf 407}, 17
\bibitem[Komissarov et al.(2009)]{KV09} Komissarov, S. S., Vlahakis,
  N., K\"onigl, A. \& Barkov, M.V. (2009), MNRAS, {\bf 394}, 1182
\bibitem[Leismann et al.(2005)]{LA05} Leismann, T., Ant\'on, L., Aloy,
  M. A., M\"uller, E., Mart\'{\i}, J. M., Miralles, J. A. \&
  Ib\'a\~nez, J. M. (2005), A\&A, {\bf 436}, 503
\bibitem[Londrillo \& Del Zanna(2000)]{LD00} Londrillo, P. \& Del Zanna
  (2000), ApJ, {\bf 530}, 508
\bibitem[Londrillo \& Del Zanna(2004)]{LD04} Londrillo, P. \& Del Zanna
  (2004), JCP, {\bf 195}, 17
\bibitem[Mignone \& Bodo(2006)]{MB06} Mignone, A. \& Bodo G. (2006), MNRAS, {\bf
  368}, 1040
\bibitem[Mignone et al.(2007)]{MB07} Mignone, A., Bodo, G., Massaglia,
  S., Matsakos, T., Tesileanu, O., Zanni, C. \& Ferrari, A. (2007),
  ApJS, {\bf 170}, 228 
\bibitem[Mignone at al.(2010)]{MR10} Mignone, A., Rossi, P., Bodo, G.,
  Ferrari, A. \& Massaglia, S. (2010), MNRAS, {\bf 402}, 7
\bibitem[Mizuno et al.(2012)]{ML12} Mizuno, Y., Lyubarsky, Y.,
  Nishikawa, K.-I. \& Hardee, P. E. (2012), ApJ, {\bf 757}, article id. 16
\bibitem[Noble et al.(2006)]{NG06} Noble, S.C., Gammie, C.F., McKinney, J.C. \& Del
               Zanna, L. (2006), ApJ, {\bf 641}, 626
\bibitem[Porth et al.(2013)]{PK13} Porth, O., Komissarov, S. S. \&
  Keppens, R. (2013), MNRAS, {\bf 431}, L48
\bibitem[Rezzolla et al.(2011)]{RG11} Rezzolla, L., Giacomazzo, B.,
  Baiotti, L., Granot, J., Kouveliotou, C. \& Aloy, M. A. (2011),
  ApJL, {\bf 732}, article id. L6
\bibitem[Ryu et al.(1998)]{RM98} Ryu, D., Miniati, F., Jones, T. W. \&
  Frank, A. (1998), ApJ, {\bf 509}, 244
\bibitem[Shibata \& Sekiguchi(2005)]{SS05} Shibata, M. \& Sekiguchi,
  Y.-I. (2005), PRD, {\bf 72}, id. 044014
\bibitem[Shu \& Osher(1988)]{SO88} Shu, C.W. \& Osher, S.J. (1988),
  JCP, {\bf 77}, 439
\bibitem[Shu \& Osher(1989)]{SO89} Shu, C.W. \& Osher, S.J. (1989),
  JCP, {\bf 83}, 32
\bibitem[T\'oth(2000)]{To00} T\'oth, G. (2000), JCP, {\bf
  161}, 605

\end{thebibliography}


\end{document}